# Optical System Design of Bionic Compound Eye with Broad Field of View

Pengcheng Su, Yu Chen*, Jiaming Zhang, Yang LI, Chao Yang

*School of Electro-Optical Engineering, Changchun University of Science and Technology, Changchun, Jilin130000, China*

**Abstract:** In nature, many common insects have compound eyes composed of many small eyes arranged on a curved retina. This kind of vision systems have many advantages, such as small size, large FOV (field of view) and high sensitivity, which have attracted extensive attention and research from world-wide researchers. It has good application prospects in military strikes and mechanical vision. In this paper, a new type of miniature compound eye system with large FOV is designed, which contains a micro-lens array and a relay system. Hexagonal micro-lens array are spliced seamlessly as a curved shell in the designed compound eye system. The intermediate curved image formed by the curved array is converted to a planar image by introducing a relay system. After combination and optimization of the micro-lens array and the relay system, the MTF values at 89.3lp/mm for each FOV within 120.5° are greater than 0.3, and the corresponding RMS spot radii less than the radius of the Airy disk, which proves the good imaging quality for the compound eye. The clear aperture of a single micro lens is 250μm with FOV 6°. After tolerance analysis, the results show the image quality still holds good enough performance and meets the requirements of the additive manufacturing process.

## 1. Introduction

Most insects in nature have compound eyes, which are their most important photosensitive parts. Compound eyes belong to multi-aperture systems and are composed of multiple sub-eyes arranged closely together and distributed on a spherical or ellipsoidal retina[1]. Schematic diagram of a compound eye and the microscopic structure of the compound eye from a bee are shown in Fig.1(a) and Fig.1(b) respectively. The latter is obtained from CT scanning after ultrathin slicing and observed with an electron microscope. The hexagonal structures and the splicing styles between sub-eyes can be seen and distinguished clearly. Based on the biological feature of compound eyes in nature, an optical system of a compound eye with such hexagonal sub-eye splicing structure is designed in this paper[2].

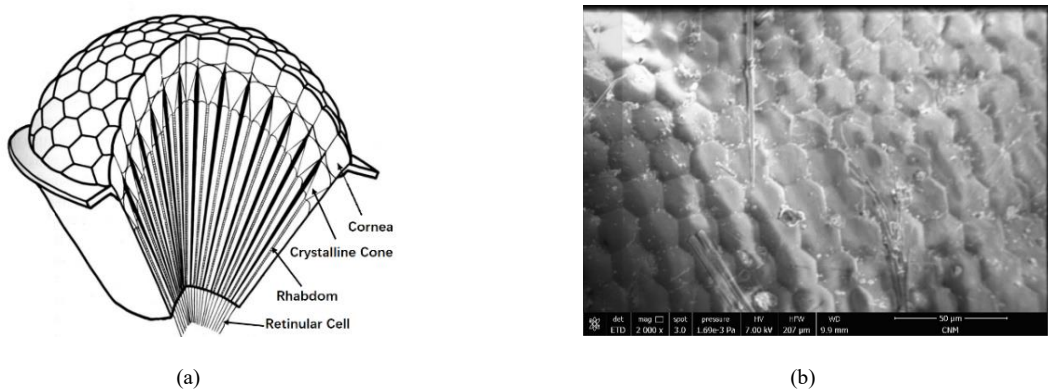

Fig.1. (a)Schematic diagram of compound eye, (b) Compound eye structure diagram of a bee

Bionic compound eye systems have developed gradually from early planar forms to current curved forms. The team leading by Allen in the United States used a three-layer optical structure to match a curved compound eye and a planar CCD. The diameter of a single micro-lens reached 500μm and the FOV of the whole system was 87°, but the alignment with high assembly precision of the three-layer structure was very complicated[3]. Shi *et al* introduced a relay system after the curved compound-eye array to convert the curved image surface to a planar detector. The FOV of the system was 122.4° and a single micro-lens was 500μm. An array of separated circular micro-lenses was adopted followed by a relay system consisting of 7 lenses. A reflective system was added to the front of a planar micro-lens array by Deng *et al* to expand the FOV of the system to 90.7°[4]. The micro-lenses were tightly spliced together with diameters of 4 mm[5]. The entire size of the compound eye system was 230 mm×107 mm×145 mm.

A miniature bionic compound eye with FOV 120.5° was designed in this paper, which consisted of a curved micro-lens array and a relay imaging system. The relay system was composed of only 4 lenses with total length 6.5 mm. The space utilization rate was improved by using a hexagonal close-splicing structure for the sub-eyes to form a spherical crown. The image quality almost achieved the diffraction limit and the desired optical parameters were also met.

*Corresponding author.
E-mail address: 323111501@qq.com

## 2. Single micro lens design and array

Design requirements of the novel bionic compound eye system are shown in Table.1.

Table. 1
Optical design requirements

| Design Parameters | Requirements |
| --- | --- |
| Full FOV of compound eye | ≥120° |
| Diameter of coronal compound eye | 5mm~10mm |
| Aperture of a single micro-lens | ≤0.3mm |
| Working waveband | F(0.486μm)~C(0.656μm) |
| Total length of the system | <10mm |

The hexagonal micro-lenses are close-spliced regularly, similar to a honeycomb structure, which spreads out from the center and forms an arrayed crown with a certain radiu[6]. Such distribution of micro-lenses can improve the space utilization rate of the compound eye effectively. All micro-lenses have the identical structural parameters and optical performance.

A complete image of the object plane can be obtained by later image processing, since a single micro-lens forms a single imaging area[6]. Thus, adjacent micro-lenses have to meet the requirements of field butting by field overlapping. The FOV of a micro-lens is generally larger than the angle between adjacent micro-lenses. But excessive overlapping of the fields of view will result in waste of system performance. Generally, FOV of the micro-lens ($\Delta\theta$) and the angle between adjacent micro-lenses ($\Delta\varphi$) are supposed to satisfy the relationship[8]:

$$\Delta\varphi < \Delta\theta < 2\Delta\varphi \qquad (2)$$

The structural schematic diagram of the micro-lens array is shown in Fig. 2.

$$\Delta\varphi = \arctan\frac{2a}{R} \qquad (3)$$

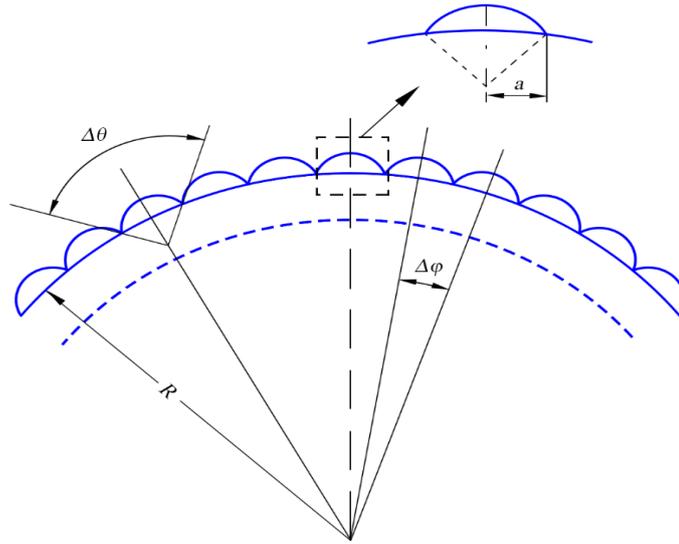

Fig.2.Structural schematic diagram of the micro-lens array

Set the radii of the front and back surfaces $r_1 = -r_2$ and $r_1 > 0$. PMMA is selected as the lens material, with refractive index 1.4918, to match the 3D printing requirements of preparation process. The corresponding optical parameters are set as follow: $f' \approx 0.93mm$ (focal length), $d \approx 0.35mm$ (micro-lens thickness) and $D = 0.5mm$ (micro-lens diameter) respectively. The formula of focal length is given as following:

$$f' = \frac{nr_1r_2}{(n-1)[n(r_2-r_1)+(n-1)d]} \qquad (1)$$

It can be solved that $r_1 = 0.97mm$ and $r_2 = -0.97mm$.

In Fig.2, R is the radius of the arrayed spherical crown. $\Delta\varphi$ is the angle between adjacent micro-lenses, and "a" is the semi-diameter of a micro-lens.

ZEMAX software is used to optimize the image quality of a single micro-lens and expand the FOV to 6° gradually. After optimization, the image distance is 0.9mm. The optimization results are presented in Fig. 3.

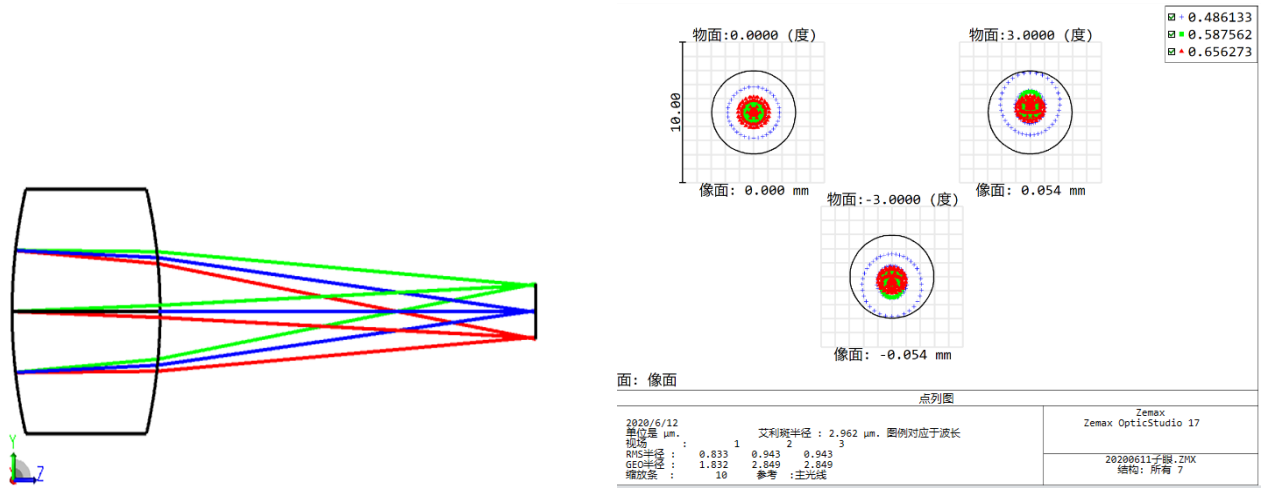

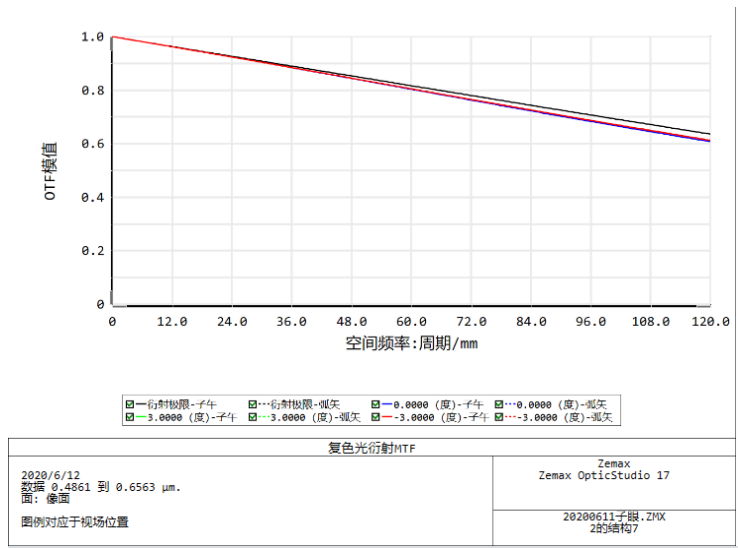

(a)

(b)

(c)

Fig. 3.(a) Layout of a single micro-lens,(b) Spot diagram,(c) MTF diagram

From Fig.3(c), it can be known that the image quality reaches the diffraction limit since the aperture and field angle are both small.

Array the designed single micro-lens to a structure of spherical crown with the radius 5mm utilizing Solidworks software, which is shown in Fig. 4[9]. The full FOV of the entire compound eye is given by formula (4).

$$\omega = k\Delta\varphi + \Delta\theta \quad (4)$$

Where, $\Delta\theta$ is the FOV of each micro-lens, $\Delta\varphi$ is the angle between two optical axes of adjacent micro-lenses, and "k" is half the number of micro-lenses in the principal section passing through the center of the spherical crown. Since $\Delta\varphi$ is 3°, Array angle is 120°. k is 40. The full FOV of the compound eye can be determined as 120.5°.

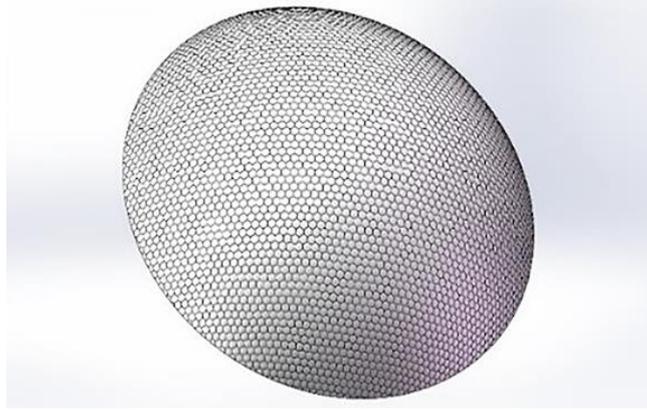

Fig. 4.3D layout of the micro-lens array

## 3. Relay system design

The curved micro-lens array forms a curved image, which cannot be received appropriately by a planar detector[10]. The imaging optical path of the micro-lenses in the principal section of the array is shown in Fig.5. Therefore, a relay system is needed to be added after the arrayed spherical crown to convert the curved image to a planar form. For a relay system of a compound eye, the FOV is supposed to be greater than 120° at least, which is also the FOV of the arrayed crown. The radius of the curved object surface for the relay system must also be consistent with the radius of the image plane for the microlens array.

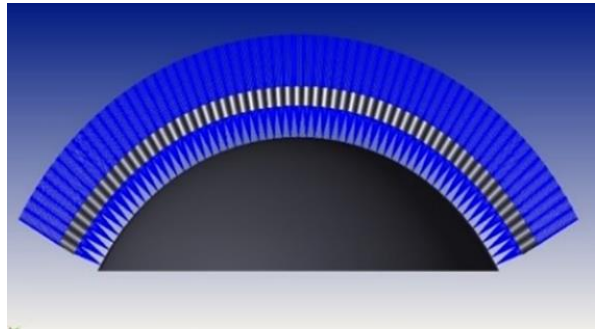

Fig.5.Imaging optical path of the micro-lens array

In this paper, an optical system in ZEBASE shown in Fig. 6 (a) is selected as the initial structure. The MTF curves of the system are given in Fig. 6(b). The total length of the system is 416mm with focal length 19.216mm. The full FOV is 180°. Current optical parameters cannot meet the requirements of design specifications[11].

8 kinds of optical glasses are adopted in the system, which all need to be changed to PMMA since the whole system is planned to be prepared by 3D printing technique. On the other side, the object is located at infinity for the initial system, which conflicts with the finite object distance of the curved intermediate image to be relayed. The initial structure is supposed to be as simplified as possible since the system aperture is greatly decreased, while keeping good enough image quality.

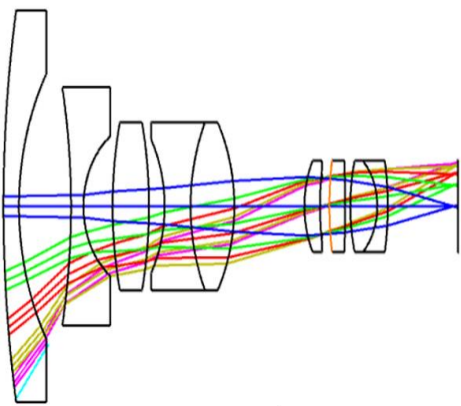
(a)

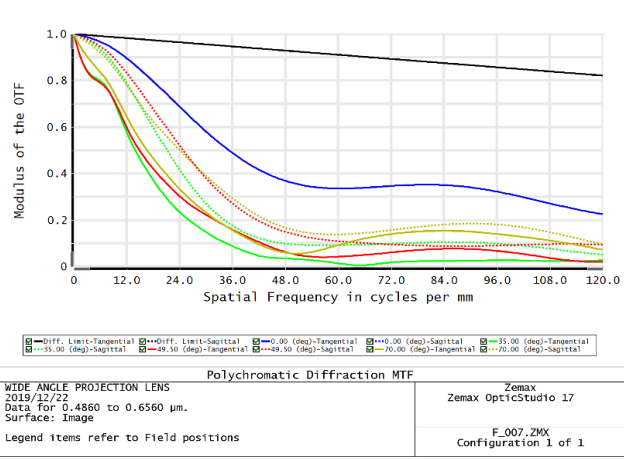
(b)

Fig. 6.(a) Structure diagram,(b) MTF diagram

The planar image is received by a 1/4-inch CCD with model SonyICX-618ALA. The diagonal length of photosensitive surface is about 4.5mm and the pixel size is 5.6μm×5.6μm. According to the law of Nyquist sampling, $N = \dfrac{1000}{2\times\alpha}$ ($\alpha$ is pixel size), the cutoff frequency can be determined as 89.3lp/mm.

Firstly, the initial system is zoomed by a factor of 1/50 considering the requirement of tube length. Optimization is carried out for the relay system after object distance is changed from infinity to finite distance. Ensure the imaging quality of the system.

Secondly, 8 different glasses are optimized to PMMA one by one by altering the refractive index and Abbe number of the materials. The cemented lens in the system is converted into a single lens in this process [12].

Finally, the conjugate distance is controlled to be less than 6.5mm while maintaining reasonable lens shapes and good enough image quality. The last two lenses are merged together by reducing their air separation to zero. Four lenses before the stop and one lens after it are removed during gradual optimization since the lenses have small optical powers and little effect to the image quality. Thus, only four lenses are retained. To balance residual aberration, the stop is moved to the front surface of the third lens.

After optimization, the layout of the relay system is shown in Fig.7(a). The MTF curve of each FOV is shown in Fig.7(b), and the spot diagram of the system is shown in Fig.7(c)

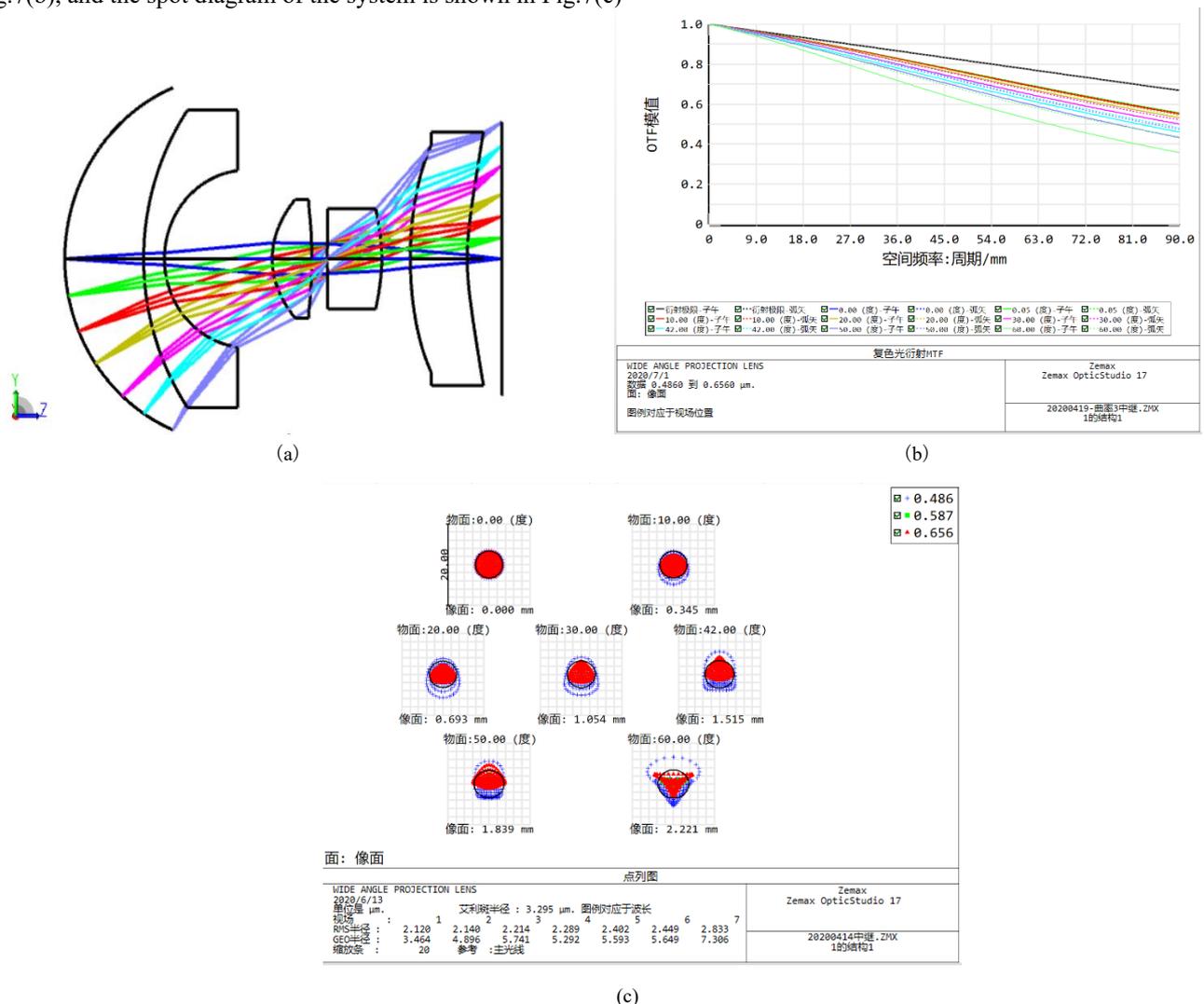

Fig. 7.(a)Layout of the relay system,(b)MTF diagram of the relay system with 120° field of view,(c)Spot diagram of the relay system

The relay system consists of four lenses with total length of 6.18 mm. MTF curves are smooth enough and MTF value of each FOV at 84lp/mm is greater than 0.5, which meets the imaging requirements.

The RMS radii of the spot diagram for different FOVs are 2.120μm, 2.140μm, 2.214μm, 2.289μm, 2.402μm, 2.449μm and 2.833μm respectively. They are all less than the radius of Airy disk 3.295μm.

## 4. Combination of micro-lens array and relay system

The micro-lens array and the relay system should be combined together to form the required compound eye system. The stop is set at the 7th surface of the combination system, which is also the same stop position of the relay system[14]. But the exit pupil of the micro-lens array does not match with the entrance pupil of the relay system, which results in degraded image quality. Since the aberration caused by the micro-lens array is small, this problem can be solved by further optimization.

The optical path diagram of seven FOVs applying multi-configuration setting are given in Fig.8. The total length of the entire compound eye system after optimization is about 9.6mm.

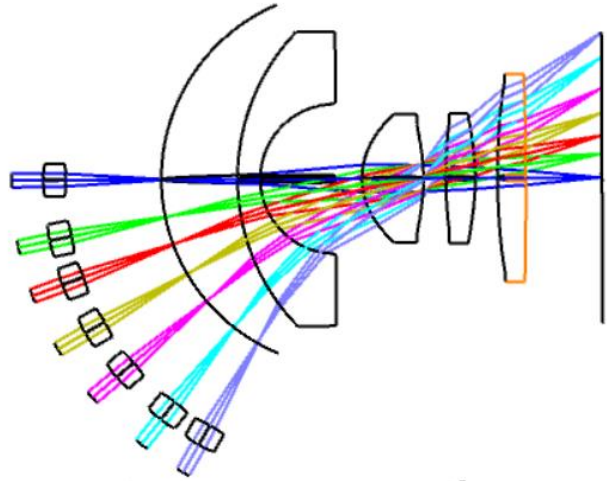

Fig.8.Optical structure of the combined eye optical system

Due to the limited space of the paper, only the MTF curves and distortion curves of three FOVs (0°, 40.07° and 60°) are shown in Fig. 9 and Fig. 10.

From Fig.9, it can be seen that MTF values of fields 0°, 40.07° and 60° are above 0.44, 0.3 and 0.29 respectively at the frequency 90lp/mm, which indicates the designed compound eye system has reached the requirements of excellent image quality.

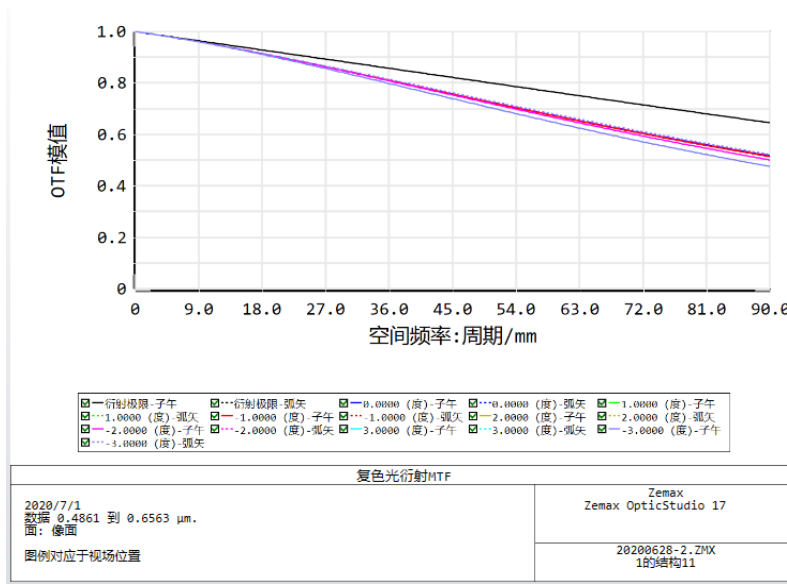

(a)

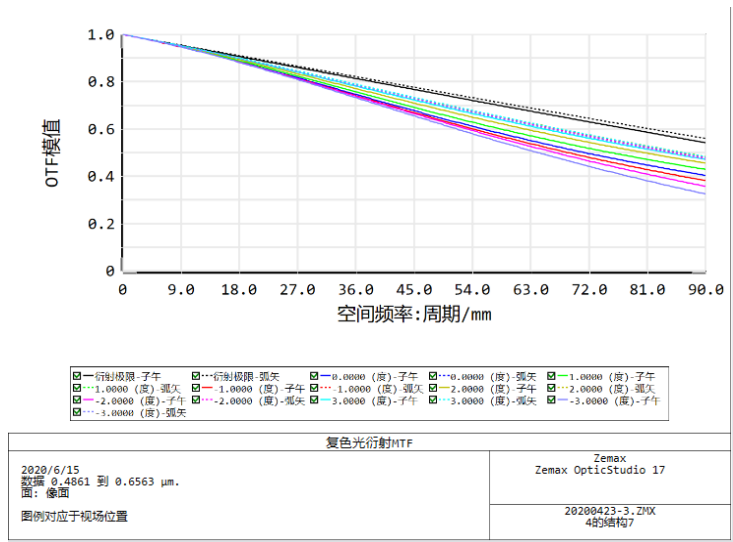

(b)

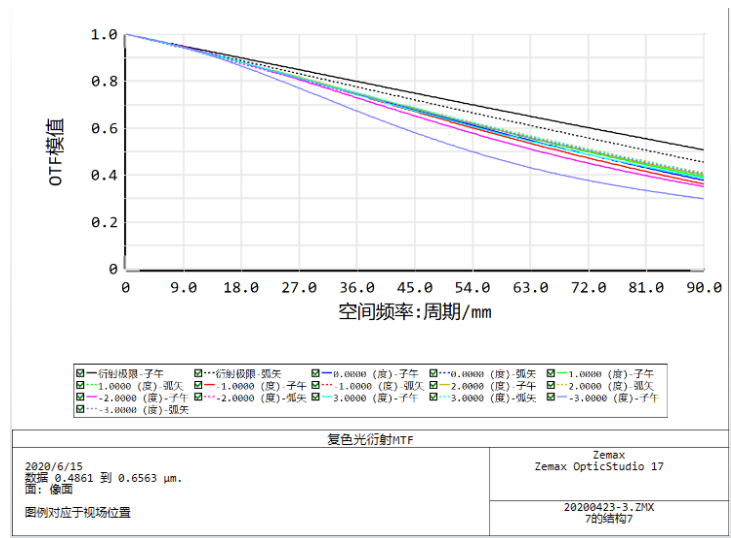

(c)

Fig. 9.(a) System MTF of field 0° (b) System MTF of field 41.38° (c) System MTF of field 60°

    The distortion increases as the field of view increases, so for compound-eye systems, the maximum distortion is determined by the microlens with a field of view of 60°. Due to space limitations, only the distortion curve of the microlens in this field of view is given below, as shown in Figure 10. It can be seen from the figure that the maximum relative distortion is less than 1%. The result of geometric image analysis indicates the letter "F" can be distinguished clearly by the designed compound eye system[15].

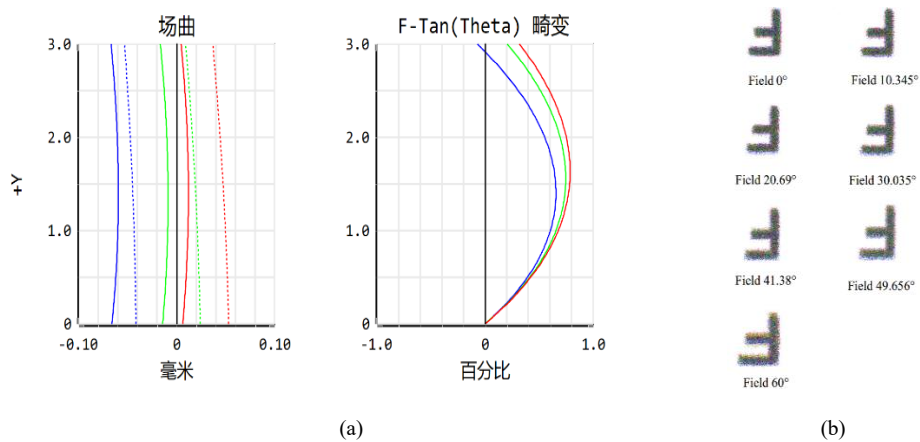

(a)                                                         (b)

Fig. 10.(a) Distortion of compound eye system at FOV of 60° (b) Result of geometric image analysis

Due to the limited space of the paper, the spot diagram of the system is shown, only the spot diagram of the system of seven FOVs are shown in Table.2.

The RMS radius of each sampling FOV is less than the radius of the Airy disk and the pixel size of the detector.

**Table.2**
Spot diagram of the compound eye system

| Config | 1 | 2 | 3 | 4 | 5 | 6 | 7 |
|---|---|---|---|---|---|---|---|
| 0° | | | | | | | |
| RMS radius (μm) | 2.283 | 2.283 | 2.396 | 2.644 | 2.701 | 2.697 | 2.939 |
| GEO radius (μm) | 3.720 | 4.687 | 5.392 | 5.262 | 5.333 | 5.865 | 7.288 |
| 3° | | | | | | | |
| RMS radius (μm) | 2.547 | 2.511 | 2.542 | 2.582 | 2.678 | 3.124 | 2.632 |
| GEO radius (μm) | 7.927 | 7.780 | 7.842 | 7.471 | 7.875 | 9.418 | 6.459 |
| -3° | | | | | | | |
| RMS radius (μm) | 2.541 | 2.729 | 2.924 | 3.102 | 3.172 | 3.354 | 4.156 |
| GEO radius (μm) | 7.927 | 9.199 | 10.110 | 10.091 | 9.949 | 9.484 | 10.687 |
| Airy radius (μm) | 4.365 | 4.371 | 4.442 | 4.617 | 4.732 | 4.972 | 5.034 |

The 3D printing equipment of model S130 is used for the preparation of the compound eye system. The supporting structure of the system is designed, in which the columnar structure can facilitate the outflow of printing liquid during the printing process and thus ensure the printing accuracy [12]. In order to avoid the negative effect of stray light on the side of the compound eye, the "lens barrel", the stop and the base support are printed integratedly with black resin material. The structure consisting of the last two lenses are printed separately as the rear fixed group. The compound eye and the first two lenses are connected by a columnar structure as the front fixed group. The connection between the lens and the lens barrel is fixed with spacers and pressure rings. Finally, the front and rear fixed groups are installed on the lens barrel in the form of snaps. The overall structure is shown in Figure 12.

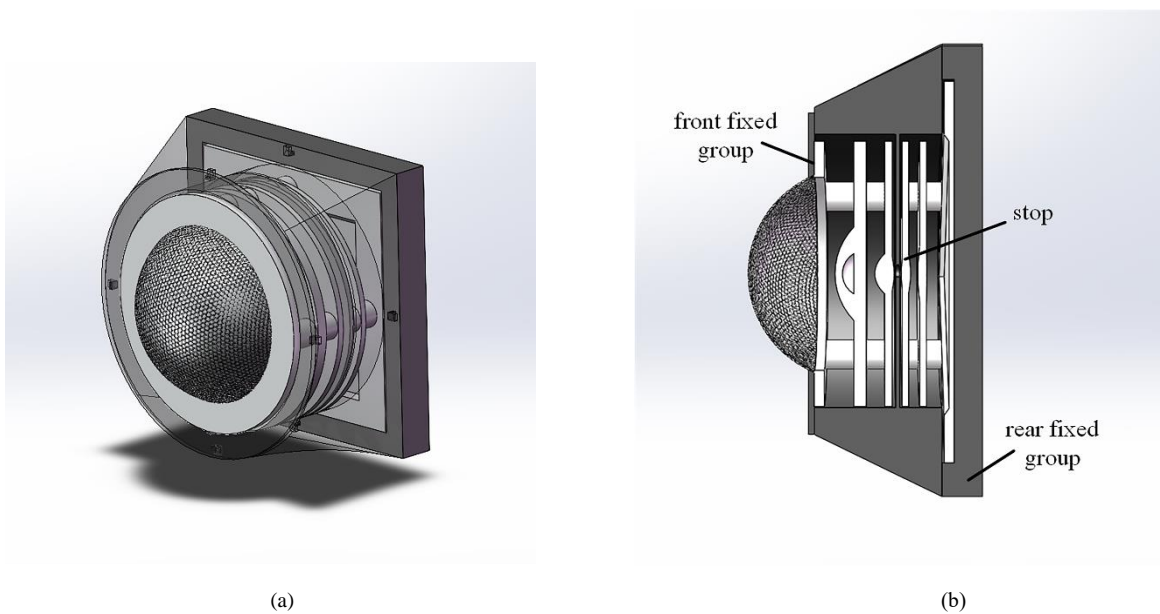

(a)　　　　　　　　　　　　　　　　　(b)

Fig.12.(a) Overall view of mechanical structure,(b) Section of Mechanical structure

In order to ensure the feasibility of the fabrication, the tolerance analysis of the entire compound eye system is required[16]. The highest printing precision of the 3D laser printer is 2μm. Therefore, the tolerance range is set to 2μm and the "MTF average" mode is selected as the imaging standard for tolerance analysis. The methods of sensitivity and Monte Carlo are utilized to analyze 1000 groups of lens data. The analysis result is shown in Table.3.

**Table.3**
Tolerance analysis result of 1000 groups of lens data

| Data | Result |
| --- | --- |
| Monte Carlo simulation times | 1000 |
| Nominal MTF value | 0.348 542 13 |
| Average MTF value | 0.347 187 12 |
| 98% lens MTF value | >0.336 334 57 |
| 80% lens MTF value | >0.343 335 32 |
| 50% lens MTF value | >0.348 132 40 |
| 20% lens MTF value | >0.351 250 35 |
| 10% lens MTF value | >0.352 350 71 |
| 2% lens MTF value | >0.353 558 39 |

It can be seen from Table.3 that the nominal value of the system MTF is 0.34854213. MTF values of 98% and 10% of the lenses are greater than 0.33633457 and 0.35235071 respectively. From above data, the change amount of MTF values after tolerance analysis is very small, which can satisfy the requirement of image quality[16].

## 5. Conclusion

An optical system of bionic compound eye with large FOV is designed. The curved image formed by the micro-lens array is converted to a planar image plane by introducing a relay system. The clear aperture of a single micro-lens is 0.25mm, and the field angle is ± 3 °. The compound eye system has a tube length of 9.6mm and covers a FOV of 120.5°, which is planned to be prepared with 3D printing technique. After tolerance analysis, the MTF values of all sampling FOVs are above 0.3 at the cut-off frequency. It indicates the imaging requirement of the designed compound eye can be met. The system has important promising applying prospects in many fields, such as mechanical vision and target recognition, with the advantages of compact structure, high sensitivity and large FOV.

## Funding

National Key Research and Development Program of China (No.2018YFB1105400).

## Declaration of Competing Interest

The authors declare no conflict of interest.